\documentclass[12pt]{article}

\usepackage{epsfig,a4wide,amsmath,amssymb,amsfonts,mathrsfs}
\usepackage{mathtools}
\usepackage{color,yfonts}
\usepackage{graphicx}
\usepackage{subcaption}
\usepackage[utf8]{inputenc}

\numberwithin{equation}{section}

\newcommand{\be}{\begin{equation}}

\newcommand{\ee}{\end{equation}}
\numberwithin{equation}{section}
\newcommand{\mytitlefont}{\fontseries{mx}\selectfont}
\DeclareMathAlphabet{\titlemath}{OT1}{cmr}{mx}{n}

\begin{document}

\begin{titlepage}

\begin{center}

~\\[2cm]

{\fontsize{20pt}{0pt} \mytitlefont  Creating a Traversable Wormhole }

~\\[0.5cm]

{\fontsize{14pt}{0pt} Gary~T.~Horowitz$^{\sharp}$, Don~Marolf$^{\sharp}$, Jorge~E.~Santos$^{\diamond}$, Diandian~Wang$^{\sharp}$}

~\\[0.1cm]

\it{ ${}^{\sharp}$  Department of Physics, University of California, Santa Barbara, CA 93106}

~\\[0.05cm]

\it{ ${}^{\diamond}$ Department of Applied Mathematics and Theoretical Physics,}
\\
\it{University of Cambridge, Wilberforce Road, Cambridge CB3 0WA, UK}

\end{center}

  \vspace{60pt}

\noindent
We argue that one can nucleate a traversable wormhole via a nonperturbative process in quantum gravity. To support this, we construct spacetimes in which there are instantons giving a finite probability for a test cosmic string  to break and produce two particles on its ends. One should be able to replace the particles with small black holes with only small changes to the spacetime away from the horizons. The black holes are then created with their horizons identified, so this is an example of nucleating a wormhole. Unlike previous examples where the created black holes accelerate apart, in our case they remain essentially at rest. This is important since wormholes become harder and harder to make traversable as their mouths become widely separated, and since traversability can be destroyed by Unruh radiation.   In our case, back-reaction from quantum fields can make the wormhole traversable.

\vfill

    \noindent

  \end{titlepage}

   \newpage

  \tableofcontents
\baselineskip=16pt
\section{Introduction}

Classically, there are no traversable wormholes. This is a consequence of topological censorship \cite{Friedman:1993ty}, which says that if  the null energy condition is satisfied, any causal curve that starts and ends at infinity can be continuously deformed to a causal curve that  stays in the asymptotic region. However, it has recently been shown that quantum matter fields can provide enough negative energy to allow some wormholes to become traversable. This was first shown for asymptotically anti-de Sitter wormholes using ideas from gauge/gravity duality \cite{Gao:2016bin, Maldacena:2017axo,Maldacena:2018lmt} and later generalized to asymptotically flat wormholes \cite{Maldacena:2018gjk,Fu:2018oaq}.\footnote{In these examples, it takes longer to go through the wormhole than to go around. As these papers explain, one expects this will be true for all traversable wormholes since it is required by the generalized second law \cite{C:2013uza} and boundary causality (when there is a field theory dual). }

While the wormhole may only be traversable for a limited time, in all of the examples that have been discussed so far the wormhole itself is eternal, i.e., it exists on all complete spacelike surfaces. This is not surprising since the quantum fields produce only small effects and there is a theorem which says that the topology of space cannot change in classical general relativity \cite{Geroch:1967fs}.

The topology of space {\it can} change through a nonperturbative quantum tunneling event. This tunneling process can be investigated in the path integral formulation of quantum gravity if a sum over all topologies is assumed in the integral. Instantons, classical solutions to Euclidean field equations, have been widely used to study  non-perturbative quantum  processes of this type \cite{ColemanChap7}. For example, there is an instanton describing the pair creation of oppositely charged black holes in a background electric or magnetic field \cite{Garfinkle:1990eq}.  This is directly analogous to the Schwinger pair creation of oppositely charged particles in an electric field. The black holes are created with their horizons identified, so this process creates a wormhole.  In the small black hole limit, the instanton reduces to the standard one-particle description of the Schwinger effect as arising from a Euclidean solution where the black hole particle moves on a circle about the origin \cite{Affleck:1981ag}. Several other analogous examples have also been found and studied \cite{Dowker:1993bt, Garfinkle:1993xk, Dowker:1994up, Hawking:1994ii, Mann:1995vb, Hawking:1995zn, Eardley:1995au}.

If the mouths of the wormhole followed static trajectories, the path integral that computes quantum corrections to the wormhole would be of the form studied in \cite{Fu:2018oaq}.  So, as argued there, in a semiclassical approximation with the metric determined by the expectation value of the quantum stress tensor, and with appropriate boundary conditions, back-reaction from quantum fields would generically transform the solution into a traversable wormhole. Justifying this approximation typically requires that many degrees of freedom propagate in the throat \cite{Maldacena:2018gjk}.
 
   Unfortunately, although the above instantons create black holes at rest, the background fields cause the black holes to quickly accelerate away from one another so that the analysis of \cite{Fu:2018oaq} does not apply.
Indeed, the black holes are separated by an acceleration horizon.
Mathematically, this allows a Killing symmetry that generates both horizons and requires any quantum stress tensor to vanish when contracted with the horizon generators.  This means that back-reaction from quantum fields will not make the wormhole traversable.  Physically, one finds that the wormhole is created in an excited state dictated by the Unruh temperature defined by the acceleration horizons.  So while the wormhole ground state may be traversable, the probability of finding such a state in the associated thermal ensemble is small due to both the tiny gap separating the energies of traversable and collapsing wormholes and the small number of traversable wormhole states relative to non-traversable black holes \cite{Maldacena:2018lmt,Maldacena:2018gjk}.  If one tries to tune parameters to reduce the acceleration, the black holes are created farther and farther apart.  Not only might one like to create wormhole mouths at finite locations, but a wide separation between the mouths makes the wormhole even more fragile so that the same problem remains.

In particular, the asymptotically flat  four-dimensional traversable wormholes of \cite{Maldacena:2018gjk} with radius $r_0$ and separation $d$ have a large parameter $q \sim r_0/\ell_{planck}$ that allows the gap $E_{gap}\sim q/d$ between traversable and collapsing wormholes to be parametrically large compared with the temperature $T=\frac{1}{2\pi d}$ set by the acceleration horizons.  But the number of traversable wormhole states is small and the black hole entropy $S_{BH} = A/4\ell_{planck}^2 =  \pi r_0^2/\ell_{planck}^2$ is much larger than $E_{gap} / T \sim r_0/\ell_{planck}$, so non-traversable wormholes dominate the ensemble at all accelerations  $a\sim 1/d$.   On the other hand, in the appropriate background such traversable wormholes will dominate if we can find instantons with
\begin{equation}
\label{eq:cMwh}
ad \ll \ell_{planck}/r_{0}.
\end{equation}
We will indeed find backgrounds that admit instantons with vanishingly small accelerations, though ours will be asymptotically anti-de Sitter.

The goal of this paper is to argue that one can  produce  traversable wormholes through a quantum tunneling event. As mentioned above, to achieve this we need to create the wormhole mouths a finite distance apart with arbitrarily small acceleration.   We can think of this as creating a pair of black holes whose horizons are initially identified, with the understanding that back-reaction from quantum fields will then render the wormhole traversable.  So long as the saddle-point black holes are near extremality and the theory contains massless fields that can propagate in the wormhole throat with appropriate boundary conditions, the wormhole will remain open for a long time \cite{Maldacena:2018lmt,Maldacena:2018gjk,Fu:2018oaq}.  Fields of this type include massless bulk fields that allow $s$-waves (so unfortunately this fails for photons or gravitons), massless modes of fermions bound to magnetic field lines, and massless excitations on cosmic strings threading the wormhole.  Under the proper boundary conditions, fields that propagate less well in the throat will still make the wormhole traversable but, if only such fields are present, the time the wormhole remains open is generally small and vanishes at extremality.

For simplicity, we will model the black holes as point particles and consider a process in which a cosmic string breaks and nucleates the two particles on its ends. We will also neglect the backreaction of the cosmic string and created particles.\footnote{For  discussions of a gravitating cosmic string nucleating two black holes which accelerate apart, see \cite{Hawking:1995zn, Eardley:1995au}.}   We will explicitly construct spacetimes in which a test string can nucleate two test particles with arbitrarily small acceleration.  In effect, we find situations in which the standard circular Schwinger instanton mentioned above is distorted into a highly eccentric configuration whose extent in the Euclidean time direction is much larger than its width in space.  One should be able to replace our particles with small black holes with negligible changes to the geometry elsewhere, and one may choose to identify their initial horizons as desired.  If the black holes have the correct temperature to match the periodicity of the particle orbit in Euclidean time, then the full instanton will be smooth.  But we may also consider instantons with conical singularities at the horizons as in \cite{Bousso:1998na}.

In Section 2, we explain a subtlety with the instantons we use. In section 3 we derive the conditions on the spacetime in order to create nearly-static particles. The calculation of the action for our instantons is given in section 4. Section 5 provides a simple example of a star in AdS which satisfies our conditions, while section 6 describes a class of vacuum examples. The latter are obtained by deforming the metric on the boundary of global AdS. Section 7 concludes the paper with some final remarks.


\section{An Instanton Subtlety}

Before proceeding to the construction of suitable spacetimes, we explain a subtlety with the instanton we need for our problem. The energy required for the nucleation of black holes (or particles) is provided by the reduction in the potential energy of the system. For example, with a magnetic field background, the appearance of a magnetic monopole pair reduces the electromagnetic potential energy; for the breaking of a cosmic string, the removal of a piece of  string reduces the elastic potential energy. Unfortunately, the same feature that provides the required energy also provides a force for the created particles to accelerate away from each other. In order to create nearly static particles, we must somehow counterbalance this undesired acceleration. In this paper, we show that the curvature of spacetime can provide this counterbalance.

One cannot create completely static particles  since if the Lorentzian solution for particles attached to cosmic strings describes two static worldlines, the Euclidean solution will also consist of two worldlines which are independent of Euclidean time. So there will not be a slice of the Euclidean solution  describing the initial unbroken cosmic string. Instead we want a family of instantons which describe the creation of particles with arbitrarily small acceleration, and thus finite but arbitrarily small Unruh temperature.

\begin{figure}[h]
    \centering
    \begin{subfigure}[b]{0.45\textwidth}
        \includegraphics[width=\textwidth]{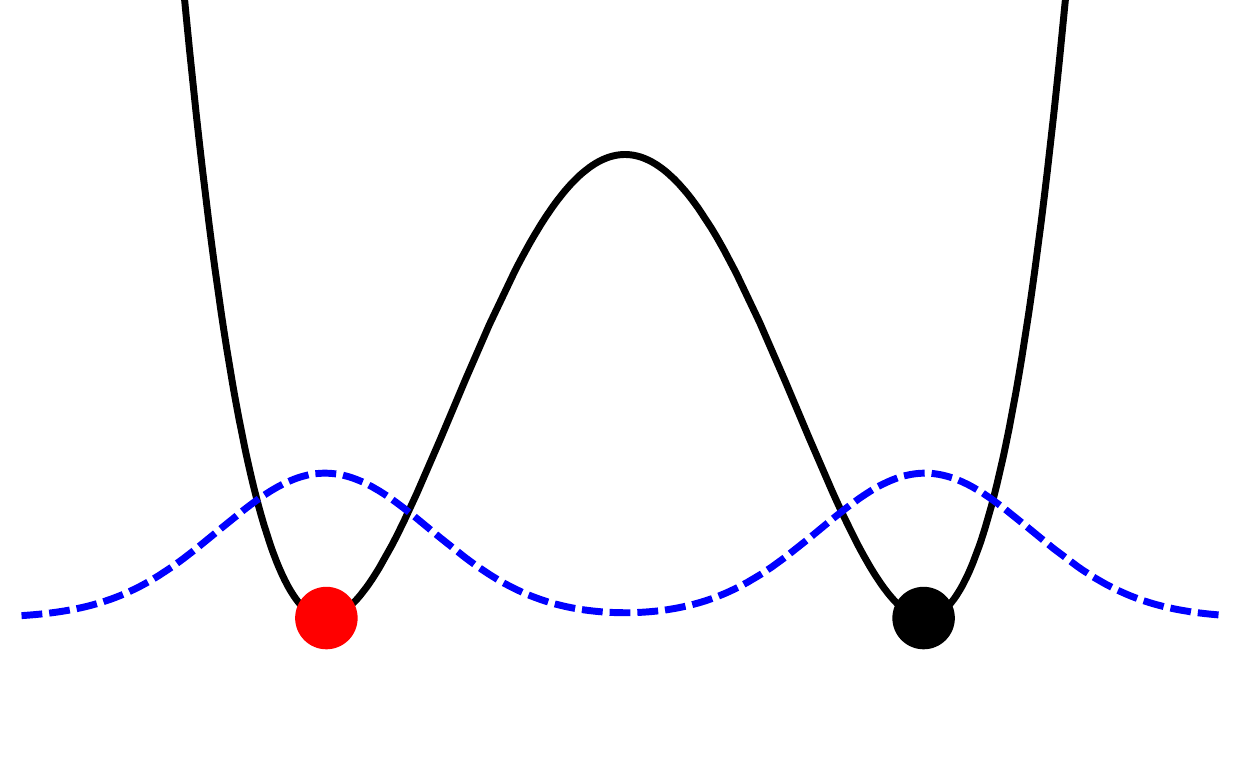}
        \caption{Degenerate minima}
        \label{fig:ground}
    \end{subfigure}
    ~ 
    \begin{subfigure}[b]{0.45\textwidth}
        \includegraphics[width=\textwidth]{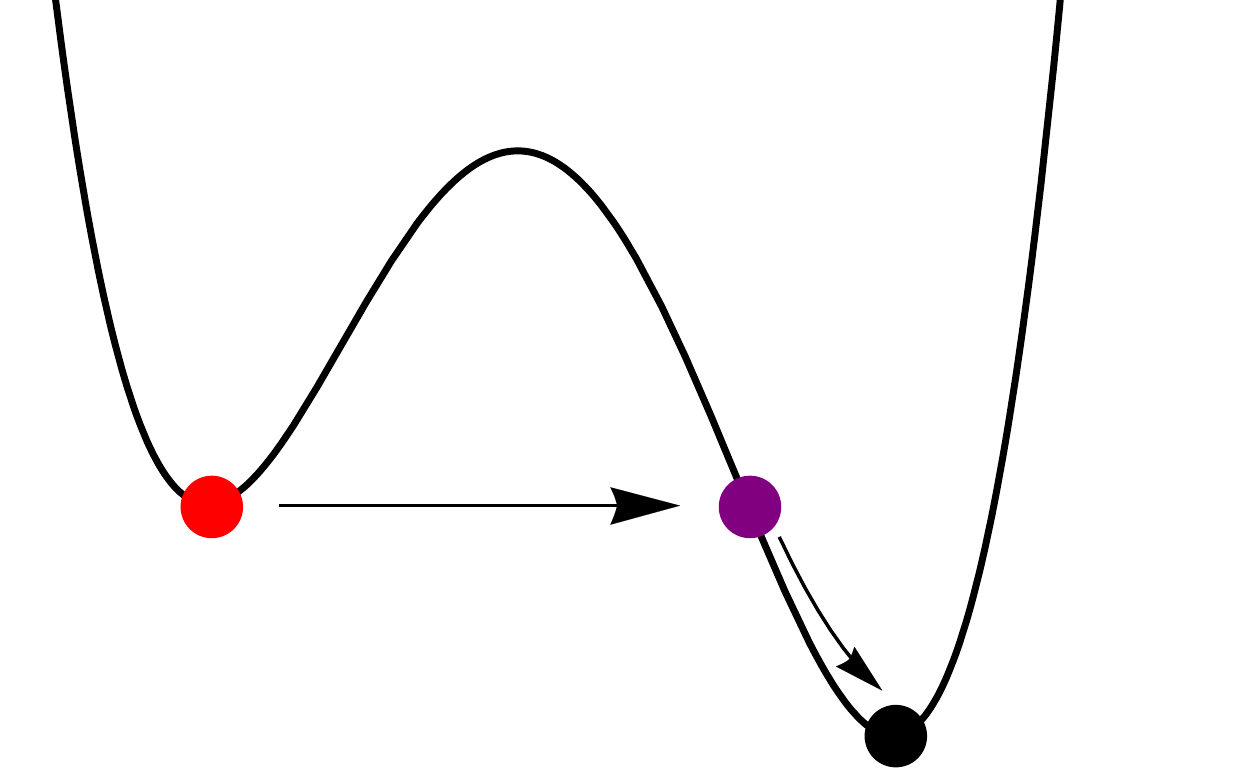}
        \caption{Nondegenerate minima}
        \label{fig:tunneling}
    \end{subfigure}
    \caption{(a) The red dot (unbroken string configuration) and black dot (broken string with static particles on the ends) are degenerate. In this case, the wavefunction of the true ground state (blue dashed curve) is a symmetrized state centered around both minima. (b) The red dot tunnels to the purple dot  (broken string with particles momentarily at rest on ends), which then oscillates around the black dot (the true vacuum). As the static pair creation limit is approached, the black dot moves up towards the purple dot so the difference in the potential energy between the red dot and the black dot diminishes.}
\end{figure}

It is instructive to think in terms of effective potentials. If the static Lorentzian solution has exactly the same energy as the unbroken cosmic string, then the effective potential has two degenerate minima (see Fig.~\ref{fig:ground}). One minimum represents the unbroken string and the other represents the broken string with static particles on its ends. There is no ``decay" from one minimum to the other in this double well potential.  However, we will see that by adjusting the ratio of the string tension  to the mass of the particles, one can lower the second minimum continuously. Now there are  instantons which describe the nucleation of two particles on the ends of the string (see Fig.~\ref{fig:tunneling}).  Assuming the original static solution is stable to small fluctuations (which will be the case in our examples) these particles will oscillate around the true minimum. We can now tune this acceleration to be as small as desired by adjusting the second minimum. The action for the instanton relative to the unbroken cosmic string, $\Delta S$, remains finite in the limit that the minima become degenerate, but the decay probability goes to zero. This is because the the decay rate is given by
\be
\Gamma = |K|e^{-2 \Delta S}
\ee
where $K$ is the contribution from small fluctuations about the instanton. This vanishes in the limit since the negative mode goes away.

In this paper, we will calculate $\Delta S$ in the limit that the minima become degenerate and interpret it as providing the leading semiclassical decay rate when the minima are slightly separate. We will also refer to the static separation as the ``separation between the created particles" with the understanding that this is the limiting case. As discussed in the introduction, our instantons will create traversable wormholes so long as the acceleration is sufficiently small, proportional to a power of $\ell_{planck}$  (see \eqref{eq:cMwh} for the case of the wormholes of \cite{Maldacena:2018gjk}).\footnote{Alternatively, one can work with the strict double well potential. Since a state localized in the first minimum is a linear superposition of the symmetric and antisymmetric energy eigenstates,  if we start with the unbroken cosmic string and wait, it will turn into a broken string with static particles on its end. At that time we can place an obstacle between the two masses to stop it from transitioning back into an unbroken cosmic string.  This is another way to create a traversable wormhole.} We expect $K$ for such cases to be proportional to some positive power of the $\ell_{planck}$, and $\Delta S$ to diverge like an inverse power of $\ell_{planck}$, so the leading contribution  to $\Gamma$ as $\ell_{planck} \rightarrow 0$ is indeed $e^{-2\Delta S}$.
\section{Conditions on the Spacetime}
We are interested in static metrics that admit a two-dimensional totally geodesic sub-manifold. Totally geodesic submanifolds have vanishing extrinsic curvature and in particular are extremal surfaces.  We may thus take this surface to be the worldsheet of our cosmic string in the case where it does not decay.   So long as no external forces act transversely to this submanifold, for appropriate initial conditions the totally geodesic condition further allows us to compute the motion of particles using only the induced two-dimensional metric
\begin{align}
\mathrm{d}s^2_{2D} = - f(r) \mathrm{d}t^2 + \frac{ \mathrm{d}r^2}{g(r)}\,.
\end{align}
In particular, this will suffice to study the motion of particles created by the decay of our cosmic string.   Furthermore, in the cases we will study below, the spacetime metric has an additional $\mathbb{Z}_2$ symmetry $r \rightarrow -r$.

\begin{figure}[h]
    \centering
    \begin{subfigure}[b]{0.45\textwidth}
        \includegraphics[width=\textwidth]{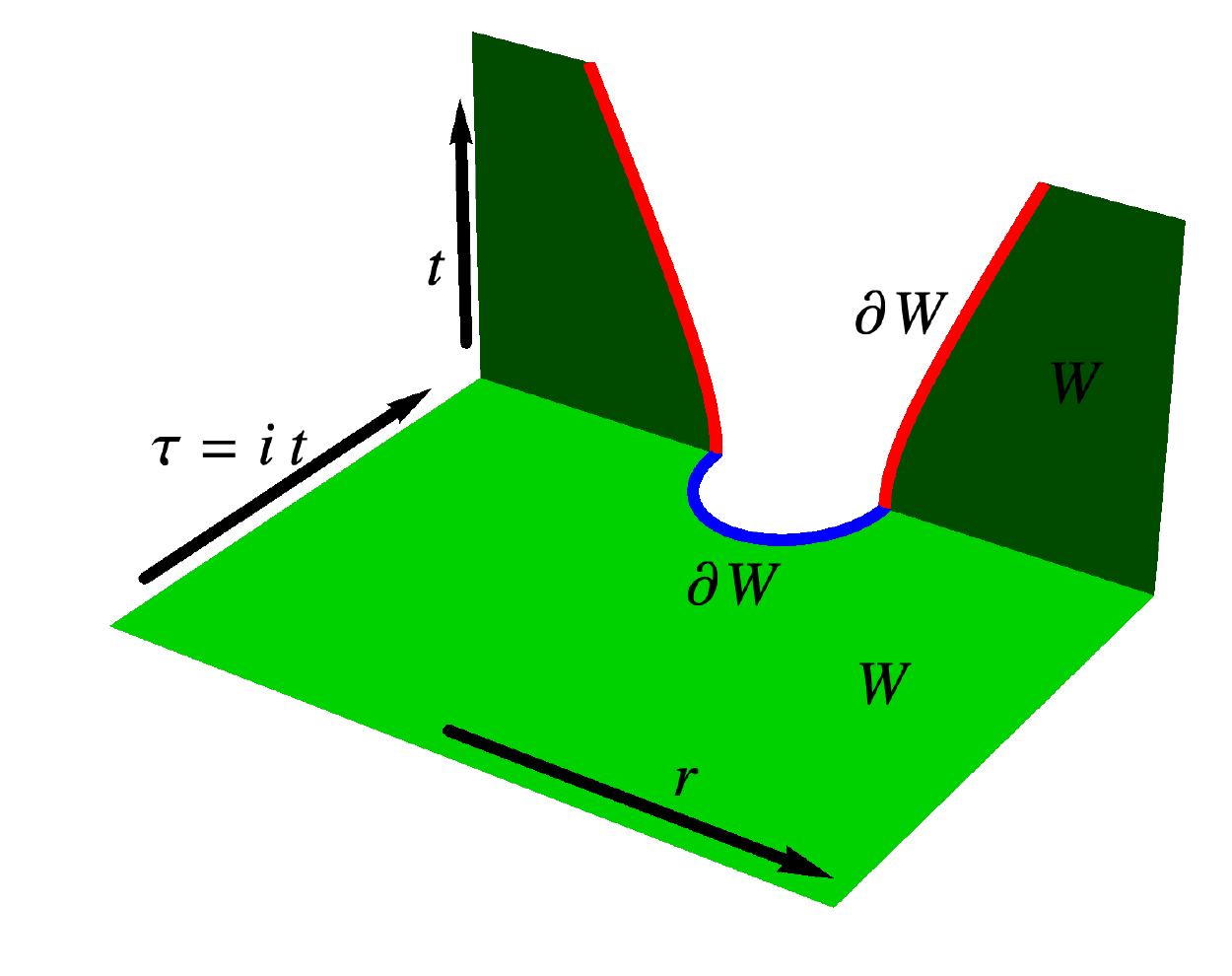}
        \caption{Subsequent acceleration}
        \label{fig:step}
    \end{subfigure}
    ~ 
    \begin{subfigure}[b]{0.45\textwidth}
        \includegraphics[width=\textwidth]{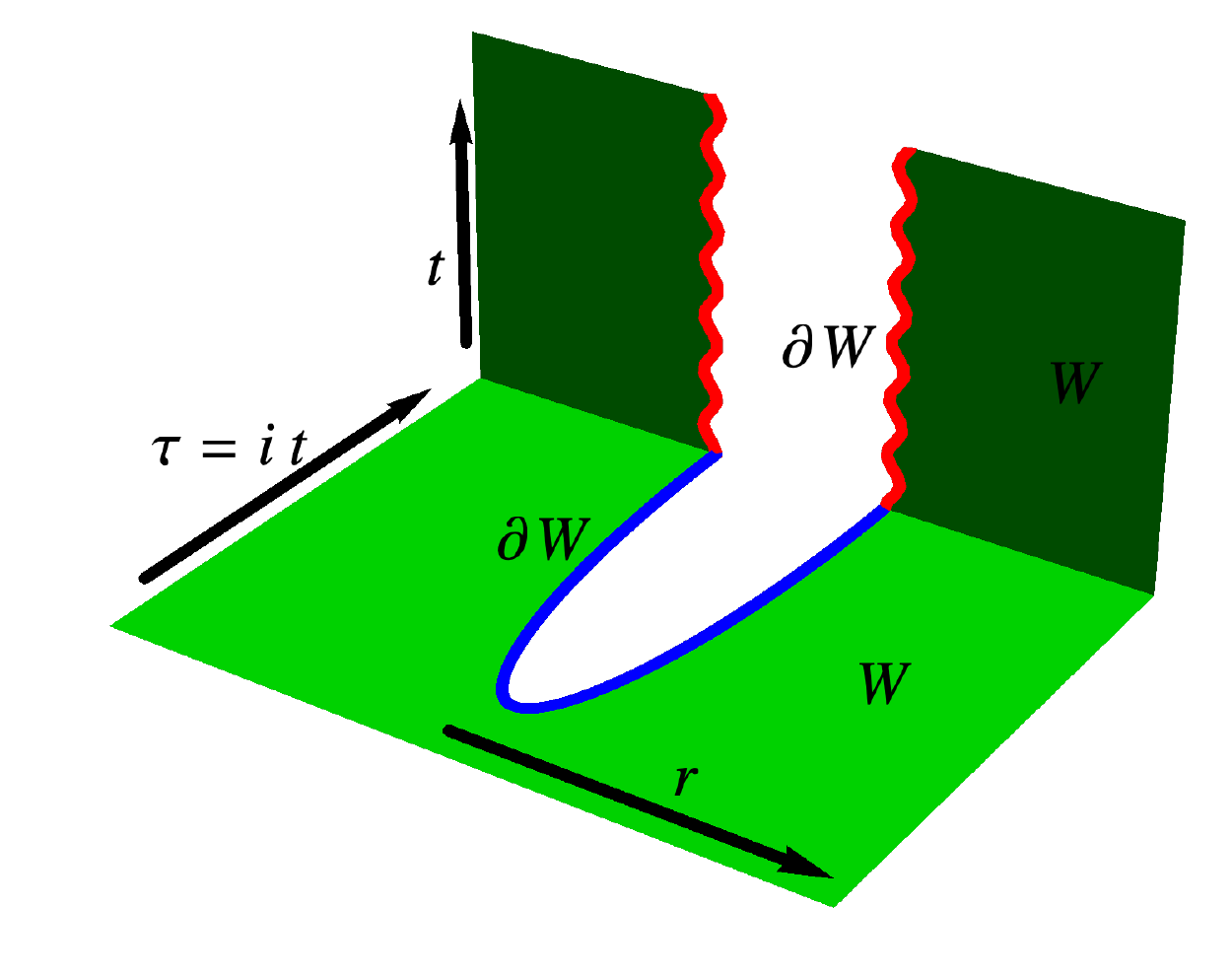}
        \caption{Subsequent oscillation}
        \label{fig:step2}
    \end{subfigure}
    \caption{The process of string breaking. The green regions in the vertical and horizontal planes are the string worldsheet $W$ in the Lorentzian and Euclidean sections respectively. The blue curve represents the Euclidean trajectory of the particle, while the red curves represent the worldlines of the physical particles after being created. (a) The traditional case where the two particles subsequently fly apart, being pulled by the broken pieces of string. (b) Our case where the particles are created close to local minima of an effective potential $V(r)$, so that the subsequent motion is oscillatory with small acceleration. Note that the Euclidean solution in this case is not a symmetric circle but is elongated so the acceleration at the transition is smaller.} \label{fig:steps}
\end{figure}

We would like to understand under what conditions one can find a nearly static particle pair created by cosmic string breaking in such backgrounds. We thus consider the following action
\begin{align}
S_L = -\mu \iint_{W} \mathrm{d}A - m \int_{\partial W} \sqrt{-g_{ab}\frac{\mathrm{d}x^a}{\mathrm{d}\lambda}\frac{\mathrm{d}x^b}{\mathrm{d}\lambda}}  \: \mathrm{d}\lambda,
\end{align}
where the subscript $L$ denotes Lorentzian signature, $W$ is the worldsheet of the unbroken string or the union of both string worldsheets that arise from breaking the original cosmic string, $\mathrm{d}A$ is the proper area element on the worldsheet,  $\lambda$ is a parameter along the boundary $\partial W$ of $W$, $\mu$ is the string tension, and $m$ is the particle mass. In our set up, $W$ has either no (inner) boundary or two boundaries, corresponding to an infinitely long string that is either unbroken or broken (see Fig.~\ref{fig:steps}).

We are interested in situations when the string is broken while preserving the $\mathbb{Z}_2$ symmetry. It is convenient to subtract the action of the unbroken cosmic string and work with the difference:
\begin{align}
\label{eq:exp}
\Delta S_L&= -2\mu \int_{-\infty}^{+\infty}\int^{r(t)}_0\sqrt{\frac{f(\tilde{r})}{g(\tilde{r})}}\: \mathrm{d}\tilde{r} \mathrm{d}t - 2m \int_{-\infty}^{+\infty} \sqrt{f(r(t))-\dfrac{\dot{r}^2(t)}{g(r(t))}} \: \mathrm{d}t,\nonumber \\
&\equiv 2\int_{-\infty}^{+\infty}\mathrm{d}t\;\mathcal{L}(r(t),\dot{r}(t)),
\end{align}
where $\dot{}$ represents derivatives with respect to $t$ and the last equality defines the Lagrangian $\mathcal{L}(r,\dot{r})$:

\begin{subequations}
\begin{equation}
\mathcal{L}(r,\dot{r})=\mu P(r)-m \sqrt{f(r)-\dfrac{\dot{r}^2}{g(r)}}\,,
\end{equation}
where
\begin{equation}
P(r)=\int_0^{r}\sqrt{\frac{f(\tilde{r})}{g(\tilde{r})}}\,\mathrm{d}\tilde{r}\geq0\,.
\end{equation}
\end{subequations}
The equations of motion for the particles derived from this action reads
\begin{equation}
\frac{\mu}{m} \sqrt{\frac{f(r)}{g(r)}} = \frac{\partial}{\partial r} \left(  \sqrt{f(r)-\frac{\dot{r}^2}{g(r)}}\right) -\frac{d}{dt}\left[ \frac{\partial}{\partial \dot{r}}\left(\sqrt{f(r)-\frac{\dot{r}^2}{g(r)}}\right)\right].
\label{eq:secondorder}
\end{equation}

Since $\mathcal{L}(r,\dot{r})$ does not depend explicitly on time, its associated Hamiltonian is conserved:
\begin{equation}
E = \dot{r}\frac{\partial \mathcal{L}}{\partial \dot{r}}-\mathcal{L}=\frac{m f}{\sqrt{f-\frac{\dot{r}^2}{g}}}-\mu P(r)\,.
\label{eq:E}
\end{equation}
This energy represents the difference between the energy of the given configuration and that of the infinite unbroken cosmic string. Since the process of nucleation conserves energy, we set $E=0$.
As a result, $\dot{r}$ is determined by the simple condition
\begin{subequations}
\begin{equation}
\dot{r}^2+V(r)=0\,,
\end{equation}
where
\begin{equation}
V(r)=f g\left\{\frac{m^2\,f}{[\mu P(r)]^2}-1\right\}\,.
\end{equation}
\label{eqs:firstorder}
\end{subequations}
Once can explicitly check that Eqs.~(\ref{eqs:firstorder}) solve Eq.~(\ref{eq:secondorder}).

The existence of static orbits is equivalent to the requirements
\begin{equation}
V(r_1)=V^\prime(r_1)=0\,.
\end{equation}
which readily give
\begin{equation}
\frac{\mu}{m}=\left. \sqrt{\frac{g}{f}}\left(\sqrt{f}\right)^\prime\right|_{r=r_1}\,,\quad\text{and}\quad \left. {f(r_1)} = {\sqrt{g}}\left(\sqrt{f}\right)^\prime P(r)\right|_{r=r_1}\,.
\label{eq:sign}
\end{equation}

We also want the static orbits to be stable.  As usual,  stability may be determined by computing $V^{\prime\prime}(r_1)$, which turns out to be given by
\begin{equation}
V^{\prime\prime}(r_1)=-\frac{g(r_1)}{f(r_1)}f^\prime(r_1)^2+\frac{1}{2}f^\prime(r_1)\,g^\prime(r_1)+g(r_1)f^{\prime\prime}(r_1)\,.
\end{equation}
In particular, if $V^{\prime\prime}(r_1)>0$, the static orbit is stable.

It is interesting to look for solutions defined by spacetimes satisfying the Einstein equations sourced by matter that respects the null energy condition. To this end, consider a static spherical metric of the form
\begin{align}
\mathrm{d}s^2_{2D} = - f(r) \mathrm{d}t^2 + \frac{ \mathrm{d}r^2}{g(r)} + r^2 \mathrm{d}\Omega_2^2\, ,
\end{align}
where $\mathrm{d}\Omega_2^2$ is the usual round metric on $S^2$,
\begin{equation}
\mathrm{d}\Omega_2^2=\mathrm{d}\theta^2+\sin^2\theta \mathrm{d}\phi^2.
\end{equation}
Consider now a null vector $u^\mu$. After expressing $u^t$ in terms of the other components of the null vector $u^\mu$, contracting $u^\mu$
 twice with the matter stress tensor $T_{\mu\nu}$ will yield
\begin{align}
8\pi T_{\mu\nu}u^\mu u^\nu = (\ldots) (u^r)^2 + (\ldots) \left[ (u^\theta)^2 +  \sin^2\theta (u^\phi)^2\right].
\end{align}
 Requiring the two terms to be non-negative then results in two conditions on the functions $f(r)$ and $g(r)$. However, when $f(r)=g(r)$, the first term always vanishes \cite{Jacobson:2007tj} and we are left with one condition on $f(r)$. In this case, the non-zero components of the Einstein tensor are
\begin{subequations}
\begin{align}
G_{tt}&=\left(\frac{-1+f+rf^\prime}{r^2}\right) (-f) \\
G_{rr}&=\left(\frac{-1+f+rf^\prime}{r^2}\right) \frac{1}{f}\\
G_{\theta\theta}&=\frac{1}{2}r(2f^\prime +rf^{\prime\prime})= \frac{G_{\phi\phi} }{\sin^2\theta}.
\end{align}
\end{subequations}
Using the Einstein equations and noting that the cosmological term vanishes when contracted with the null vector gives
\begin{align}
8\pi T_{\mu\nu}u^\mu u^\nu =\left( 1-f+\frac{1}{2}r^2 f^{\prime\prime}\right) \left[ (u^\theta)^2 +  \sin^2\theta (u^\phi)^2\right].
\end{align}
The null energy condition then requires
\begin{align}
\label{eqn:necff}
 f^{\prime\prime} \geq \frac{2(f-1)}{r^2}
\end{align}
for all $r$.

It would also be reasonable to ask that the stronger weak energy condition holds: $T_{\mu\nu}u^\mu u^\nu \geq 0$ for all timelike observers $u^\mu$. This gives an additional condition which for $f(r)=g(r)$ takes the form:
\begin{align}
r f' \leq 1+\frac{3r^2}{L^2}-f.
\end{align}

We will now use what we learned in this section to construct background geometries that admit static orbits resulting from the breaking of a cosmic string.
\section{Semiclassical Production Rate}
Besides finding orbits in a plethora of spacetimes, we are also interested in studying the instantons that give a finite probability for a test cosmic string  to break and produce particles attached to the broken ends. To do so we must solve the Euclidean equations of motion and find a trajectory $r(\tau)$ describing expansion from $r=0$ to (near) the static orbit $r=r_1$ determined in the previous section. Here $\tau$ is the Wick rotated time $\tau=it\,$. We use Euclidean time translation symmetry to set $r=r_1$ at $\tau=0$ and $r=0$ at $\tau = \tau_0 < 0$. In addition, regularity under $r\rightarrow -r$ demands $\mathrm{d}r/\mathrm{d}\tau|_{\tau=\tau_0}=+\infty\Rightarrow \tau^\prime(0)=0$ (see the blue curves in Fig.~\ref{fig:steps}).

In the semiclassical approximation, the production rate of such instantons is given by
\begin{subequations}
\begin{align}
\Gamma & = e^{-2 \Delta S}
\\
\Delta S & =  - \mu  \iint_{\overline{W}} \sqrt{\frac{f(r)}{g(r)}}\mathrm{d}r \mathrm{d}\tau+ m\int_{\partial \overline{W}} \sqrt{f(r)+\frac{1}{g(r)} \left( \frac{\mathrm{d}r}{\mathrm{d}\tau} \right)^2}  \mathrm{d}\tau \nonumber
\\
& = -2 \int_{\tau_0}^{0}\left[\mu P(r(\tau))-m \sqrt{f(r(\tau))+\frac{1}{g(r(\tau))} \left( \frac{\mathrm{d}r}{\mathrm{d}\tau} \right)^2}\right]\,\mathrm{d}\tau\,.
\end{align}
\end{subequations}
where $\Delta S$ is the Euclidean action of the instanton relative to the eternal unbroken cosmic string in the background spacetime with $S$ defined to be the analytically continued value of the Lorentzian action multiplied by $-i$, and $\overline{W}$ is the complement of the Euclidean string worldsheet.

The Euclidean equations of motion for the particles can be readily obtained by varying $\Delta S$ with respect to $r(\tau)$. Just as for Lorentzian signature, these instantons can be reduced to studying a single ODE via ``energy conservation'', which in this case gives
\begin{subequations}
\begin{equation}
\frac{\mathrm{d}r}{\mathrm{d}\tau} =\sqrt{f(r) g(r)}\sqrt{\frac{f(r)}{\left[C_1+\frac{\mu}{m}P(r)\right]^2}-1}\,,
\end{equation}
where $C_1$ parametrises the ``Euclidean energy''. Regularity of $\tau(r)$ at $r=0$ demands $C_1=0$. We thus have
\begin{equation}
\frac{\mathrm{d}r}{\mathrm{d}\tau} = \sqrt{f(r) g(r)}\sqrt{\frac{f(r)}{\frac{\mu^2}{m^2}P(r)^2}-1}\,.
\end{equation}
\end{subequations}
Inputting the expression for $\mathrm{d}r(\tau)/\mathrm{d}\tau$ into the on shell action gives a simple expression for $\Delta S$
\begin{equation}
\Delta S=2\int_{\tau_0}^0\left[-\mu P(r(\tau))+\frac{m^2}{\mu}\frac{f(r(\tau))}{P(r(\tau))}\right]\mathrm{d}\tau.
\end{equation}
Since this expression only depends on $\tau$, through $r(\tau)$, we can perform another change of variable and write everything in terms of $r$ instead. This simplifies our final expression to
\begin{equation}
\Delta S=2m\int_0^{r_1}\frac{1}{\sqrt{g(r)}}\sqrt{1-\frac{\mu^2}{m^2}\frac{P(r)^2}{f(r)}}\mathrm{d}r\,.
\label{eq:actionproduction}
\end{equation}
This final expression bypasses the need to actually determine the instanton orbits $r(\tau)$, since it depends on $f$ and $g$ only. Note that $r_1$ and $\mu/m$ are determined by  Eq.~(\ref{eq:sign}).

This is the expression we are going to use in the following sections.  For finite $r_1$, it is clear that $\Delta S$ is finite.  This may at first seem surprising as in the limit where the created particles sit on static worldlines $r = r_1$ the Euclidean time $\tau_0$ at which $r=0$ must diverge to $-\infty$.  This is the phenomenon described in the introduction where the Euclidean worldline becomes a highly eccentric closed loop that extends much farther in Euclidean time than it does in space.  In this limit, the action $\Delta S$ receives contributions from an infinite set of Euclidean times.  However, the contribution from Euclidean times near $\tau =0$ vanish in this limit as any static orbit has Euclidean action $E \int  d\tau$, so energy conservation guarantees such contributions to precisely cancel analogous contributions to the action of the unbroken cosmic string.  The net result is that $\Delta S$ receives non-negligible contributions only from Euclidean times near $\pm \tau_0$ and thus naturally remains finite as $\tau_0 \rightarrow -\infty$.

\section{An AdS Star Example}
In this section  we present a simple analytic $3+1$-dimensional metric satisfying all the conditions derived in Sec. 3. In other words, it contains  stable static solutions for particles attached to a cosmic string which have the same energy as the unbroken string. The metric is:
\begin{subequations}
\begin{align}
\mathrm{d}s^2 &= - f(r) \mathrm{d}t^2 + \frac{\mathrm{d}r^2}{f(r)} +  r^2 \mathrm{d}\Omega^2_2,&
\\
f(r) &=
\begin{cases}
\displaystyle 1+\frac{r^2}{L^2}-\frac{2M}{r},  &r > R
\\
\displaystyle 1+ \frac{r^2}{L^2} + A r^6 - B r^2 , &r<R,
\end{cases}
\end{align}
\end{subequations}
where $A$ and $B$ are positive parameters to be determined, and $R$ is for now an arbitrary parameter bigger than the horizon radius for the given $M$.

The null energy condition is satisfied everywhere, which can be checked using Eq.~(\ref{eqn:necff}). Matching $f(r)$ and $f'(r)$ at $R$ gives
\begin{align}
A &= \frac{3M}{2R^7}\\
B &= \frac{7M}{2R^3}.
\end{align}

The static and energy conservation conditions require the radius of static particles $r_1$ to be
\begin{align}
r_1 = \left(\frac{1}{2A}\right) = R \left( \frac{R}{3M}\right)^{1/6},
\end{align}
with $r_1 < R$. Thus, $R<3M$, \emph{i.e.} the star has to be fairly compact.

If we want the weak energy condition to hold, we also need $T_{00}\geq 0$. We find:
\begin{equation}
8 \pi T_{00} =\frac{21 M (R^4 - r^4)}{4 R^{14}}\left[2 \left(1 + \frac{r^2}{L^2}\right) R^7 + M r^2(3 r^4 - 7 R^4)\right]\,.
\end{equation}
There is only one term that could be negative, and we can ensure the expression is always positive by taking $M\leq \frac{2R^3}{7\,L^2}$. Note that this does not conflict with the previous inequality $3M>R$. In fact, combining these inequalities yields $(R/L)^2 > 7/6$, so the star must be larger than the AdS radius. Interestingly, note that $T_{00}$ always vanishes as $r\rightarrow R$, so that the energy density on the edge of the star is zero. So, in a sense, it is more like a compact spherical cloud than a compact spherical star. It is also interesting to note that the weak energy condition  fails if $f(r) = 1+r^2/L^2 + Ar^4-Br^2$ (for any allowed $R$ and $M$). This is why we chose an $r^6$ term in our metric.

One can explicitly compute the semi-classical production rate in this example. If we define $x$ and $y$ by
\begin{equation}
R \equiv 3\,y\,M\quad\text{and}\quad M \equiv \frac{2}{7}\frac{R^3}{L^2}\,x\, ,
\end{equation}
then the instanton action   is  given by
\begin{subequations}
\begin{multline}
\Delta S = \sqrt{\frac{7}{3}}\frac{1}{\sqrt{x} y^{1/3}}\frac{1}{(1-y_c+y_c^2)}\Bigg\{\sqrt{2-y_c} \sqrt{y_c} \left[\frac{3}{2} \arctan\left(\sqrt{\frac{2-y_c}{3y_c}}\right)-\frac{\pi }{4}\right]+\\
\frac{3 y_c^2}{2 \sqrt{4+2y_c+y_c^2}}\mathrm{arctanh}\left(\frac{\sqrt{3} \sqrt{4+2 y_c+y_c^2}}{4+y_c}\right)\Bigg\}
\end{multline}
where
\begin{equation}
y_c = \frac{2 \sqrt{7} \sqrt{1-x}}{3 \sqrt{x} y^{1/3}} \sinh \left\{\frac{1}{3} \mathrm{arcsinh}\left[\frac{27 x^{3/2} y}{7 \sqrt{7} (1-x)^{3/2}}\right]\right\}.
\end{equation}
In particular, the expression is finite as expected.

\end{subequations}
\section{Vacuum Examples}
We now construct a class of example spacetimes sourced only by a cosmological constant for which static, stable orbits can be found, and we compute the associated instanton production rate. Global AdS does not satisfy our condition (\ref{eq:sign}) since the $E=0$ condition would require the particles to be created infinitely far apart. But one can view this as being close to satisfying the condition, and hope that a small perturbation will allow the particles to be created at a finite separation.  We will show that this is indeed the case.

We construct static, axisymmetric solutions for which the boundary metric is not exactly the Einstein static universe. We seek bulk geometries for which the boundary metric takes the following simple form
\begin{equation}
\mathrm{d}s^2_{\partial} = -\mathrm{d}t^2+L^2\left[1+\epsilon\,\mathbb{S}_{\ell}(\theta)\right]\mathrm{d}\Omega_2^2
\end{equation}
where $\mathbb{S}_{\ell}(\theta)$ is a scalar harmonic which, under our symmetry assumptions, is related to the Legendre polynomials $P_{\ell}$ via $S_{\ell}(\theta)=P_{\ell}(\cos \theta)$.
To proceed, we will use a mixture of analytical and numerical techniques, each of which is explained in a separate subsection.
\subsection{Nonlinear perturbative approach}
In this subsection we will take the so called quasi-spherical gauge, where our bulk spacetime metric takes the following form
\begin{equation}
\mathrm{d}s^2 = -F(r,\theta)\left(1+\frac{r^2}{L^2}\right)\mathrm{d}t^2+\frac{\mathrm{d}r^2}{\displaystyle G(r,\theta)\left(1+\frac{r^2}{L^2}\right)}+S(r,\theta)r^2\mathrm{d}\Omega_2^2\,,
\end{equation}
where $F$, $G$ and $S$ are to be determined in what follows. For $F=G=S=1$ we recover pure AdS written in standard global coordinates. We seek solutions for which
\begin{equation}
\lim_{r \to \infty} F=\lim_{r \to \infty} G=1\,,\quad\text{and}\quad \lim_{r \to \infty} S=1+\epsilon\,\mathbb{S}_{\ell}(\theta)\,.
\end{equation}
We will demand regularity at the poles of the deformed $S^2$, located at $\theta=0,\pi$ and at the centre we impose that the geometry is smooth. This in turn implies
\begin{equation}
\frac{1}{G(\theta,0)}=S(\theta,0)\,,\quad \text{and}\qquad \left.\frac{\partial G}{\partial r}\right|_{r=0}=\left.\frac{\partial F}{\partial r}\right|_{r=0}=\left.\frac{\partial S}{\partial r}\right|_{r=0}=0\,.
\end{equation}

Since we are interested in this section in a perturbative expansion in $\epsilon$, we shall expand all our functions in terms of tensor derived spherical harmonics (see for instance \cite{Kodama:2003jz}). Given that we are working in four spacetime dimensions, there are no tensor harmonics, so we only have the scalar derived and vector derived tensor perturbations. The geometries we seek are axisymmetric with respect to $\partial/\partial \phi$ and it is a simple exercise to show that no vector harmonics on the two-sphere exists within such symmetries. We are thus left with scalar derived tensor perturbations. For a generic gauge, and imposing staticity, these are given by
\begin{equation}
\delta g_{tt}^{\ell}=f_{tt}(r) \mathbb{S}_{\ell}\,,\quad \delta g_{rr}^{\ell}=f_{rr}(r) \mathbb{S}_{\ell}\,, \quad \delta g_{ij}^{\ell}=H_L(r) \mathbb{S}_{\ell} \mathbb{G}_{ij}+H_T(r) \mathbb{S}^{\ell}_{ij}\,, \quad \text{and}\quad \delta g_{r\theta}^{\ell}=f_r(r)\nabla\!\!\!\!\nabla_\theta \mathbb{S}_{\ell}\,,
\end{equation}
where $i,j$ are indices on the two-sphere, $\nabla\!\!\!\!\nabla$ is the standard Levi-Civita connection on $S^2$, $\mathbb{G}$ is the round metric on the two-sphere and $\mathbb{S}_{ij}$ is a traceless transverse tensor with respect to $\mathbb{G}$, given by
\begin{equation}
\mathbb{S}^{\ell}_{ij}= \nabla\!\!\!\!\nabla_i \nabla\!\!\!\!\nabla_j \mathbb{S}_{\ell}+\frac{\ell(\ell+1)}{2}\,\mathbb{S}_{\ell}\,\mathbb{G}_{ij}\,.
\end{equation}
Any linear diffeomorphism, say $\xi$, can equally be decomposed in terms of such scalar harmonics
\begin{equation}
\xi^{\ell} = \xi_r(r)\mathbb{S}_{\ell}\mathrm{d}r+K(r)(\nabla\!\!\!\!\nabla_\theta \mathbb{S}_{\ell})\mathrm{d}\theta\,.
\end{equation}
Our gauge freedom amounts to choosing $\xi_{r}$ and $K(r)$ so that $H_T(r)=f_r(r)=0$. It is a simple exercise to show that this gauge choice is possible to achieve, by looking at the relevant components of the gauge transformed metric perturbation
\begin{equation}
\hat{\delta g}^{\ell}_{ab}=\delta g^{\ell}_{ab}+\nabla_{(a}\xi^{\ell}_{b)}\,.
\end{equation}

This means, in our gauge, the metric functions can be expanded as
\begin{subequations}
\begin{align}
F(r,\theta)=1+\sum_{i=1}^{+\infty} \sum_{\ell=0}^{\infty }\epsilon^i f^{(i)}_\ell(r)\mathbb{S}_{\ell}(\theta)\,,
\\
G(r,\theta)=1+\sum_{i=1}^{+\infty} \sum_{\ell=0}^{\infty }\epsilon^i g^{(i)}_\ell(r)\mathbb{S}_{\ell}(\theta)\,,
\\
S(r,\theta)=1+\sum_{i=1}^{+\infty} \sum_{\ell=0}^{\infty }\epsilon^i s^{(i)}_\ell(r)\mathbb{S}_{\ell}(\theta)\,.
\end{align}
\end{subequations}

We shall present generic results for arbitrary values for even $\ell$ to first order in $\epsilon$, and will specialise to the higher order case for $\ell=2$. To linear order in $\epsilon$, we find
\begin{subequations}
\begin{equation}
f^{(1)}_{\ell}=g^{(1)}_{\ell}=s^{(1)}_{\ell}=0\,,\quad\text{for}\quad \ell=0 \ {\rm and} \  \ell\geq3\,,
\end{equation}
and
\begin{align}
g^{(1)}_{2}(r)&=f^{(1)}_{2}(r)\,,
\\
f^{(1)}_{2}(r)&=-\frac{(\ell +2) \Gamma \left(\frac{\ell +3}{2}\right)^2}{\sqrt{\pi } (\ell+1) \Gamma \left(\ell +\frac{3}{2}\right)}\frac{1}{\displaystyle 1+\frac{r^2}{L^2}}\left(\frac{r}{L}\right)^{\ell } \, _2F_1\left(\frac{\ell -1}{2},\frac{\ell }{2};\ell +\frac{3}{2};-\frac{r^2}{L^2}\right)\,,
\\
s^{(1)}_{2}(r)&=-\frac{2 r^3 f_2'(r)}{L^2 \left(\ell ^2+\ell -2\right)}-\left(1 +\frac{\ell ^2+\ell -4}{\ell ^2+\ell -2}\frac{r^2}{L^2}+\frac{2}{\ell ^2+\ell -2}\frac{r^4}{L^4}\right)\frac{f_2^{(1)}(r) }{\displaystyle 1+\frac{r^2}{L^2}}\,,
\end{align}
\end{subequations}
where $\, _2F_1(a,b;c;z)$ is the Gauss hypergeometric function. The normalisation was chosen so that
\begin{equation}
\lim_{r\to+\infty}s^{(1)}_{2}(r)=1\,.
\end{equation}

We are now ready to use Eq.~(\ref{eq:sign}) to linear order in $\epsilon$. Along $\theta=0,\pi$ and for even $\ell$, we have
\begin{subequations}
\begin{align}
&f(r)=\left[1+\epsilon f^{(1)}_\ell(r)\right]\left(1+\frac{r^2}{L^2}\right)+\mathcal{O}(\epsilon^2)\,,
\\
&g(r)=\left[1+\epsilon f^{(1)}_\ell(r)\right]\left(1+\frac{r^2}{L^2}\right)+\mathcal{O}(\epsilon^2)\,.
\end{align}
\end{subequations}
which gives
\begin{equation}
P(r)=r+\mathcal{O}(\epsilon^2)\,.
\end{equation}
Once the dust settles, one finds for the critical location of the orbit:
\begin{equation}
r_1 = \frac{2 \ell }{\ell +1}\frac{\Gamma \left(\frac{\ell }{2}\right)^2}{ \Gamma \left(\frac{\ell +1}{2}\right)^2}\,\frac{L}{\epsilon}+\mathcal{O}(\epsilon^0)
\end{equation}

We will want to compare this result to our full nonlinear numerical results of the next section. In order to perform such comparison, we need a gauge invariant measure for the distance between the two orbits. This is best done by looking at the proper distance between the two orbits, given by
\begin{equation}
\mathcal{P}_{\ell}=2\int_0^{r_1}\frac{1}{\sqrt{G(r,0)}\sqrt{1+\frac{r^2}{L^2}}}\mathrm{d}r=2 L\,\log\left(\frac{2\,r_1}{L}\right)+\mathcal{O}(\epsilon)
\label{eq:prediPell}
\end{equation}
The calculation of the on-shell action is very simple  to this order in $\epsilon$, giving
\begin{equation}
\Delta S_{\ell} = \pi\,\,m\,L+\mathcal{O}(\epsilon \log\epsilon)
\label{eq:missing}
\end{equation}
with the next order term being possible to determine only if $F$, $G$ and $S$ are known to second order in $\epsilon$.

For particular values of $\ell$ we can do substantially better. For instance, we took $\ell=2$, and solved the resulting equations up to second order $\mathcal{O}(\epsilon^2)$ and thus determined the order $\mathcal{O}(\epsilon \log \epsilon)$ term missing in Eq.~(\ref{eq:missing}). The intermediate expressions are too cumbersome to be reproduced here, but follow {\it mutatis mutandis} calculations done elsewhere \cite{Costa:2015gol,Costa:2017tug,Markeviciute:2017jcp}.

For the proper distance we now find
\begin{equation}
\mathcal{P}_{\ell=2}=2 L \log \left(\frac{32}{3 \pi\,\epsilon}\right)+\frac{\left(384 C+3424-45 \pi ^2\right) L \epsilon }{512}+\mathcal{O}\left(\epsilon ^2\right)\,,
\label{eq:properpredi}
\end{equation}
where $C$ is Catalan's constant. For the instanton production rate we find
\begin{equation}
\Delta S_{\ell=2} = \pi  L m \left\{1+\frac{3}{16} \epsilon  \left[1+\frac{\pi }{2}+2 \log \left(\frac{3 \pi  \epsilon }{32}\right)\right]\right\}+\mathcal{O}\left(\epsilon ^2\log\epsilon\right)\,.
\label{eq:actionprodpre}
\end{equation}
Computing $\Delta S_{\ell=2}$ to this order in $\epsilon$ requires using the following Feynman trick twice:
\begin{equation}
\arctan(a \, \hat{x})=\int_0^{1}\frac{a \hat{x}}{1+a^2 \hat{y} ^2 \hat{x}^2}\mathrm{d}\hat{y}\,\quad\text{where}\quad a,\hat{x}\in\mathbb{R}\,.
\end{equation}

\subsection{Exact numerical construction}
We now move beyond perturbation theory and numerically construct exact vacuum solutions with deformed boundary metrics. We will use  the DeTurck formalism, which was first proposed in \cite{Headrick:2009pv} and reviewed in \cite{Wiseman:2011by,Dias:2015nua}. Just as the preceding sections, we work in four bulk spacetime dimensions, but our methods work equally well in higher dimensions\footnote{In odd spacetime dimensions we would have poor convergence properties because of conformal anomalies.}.

The starting point of the DeTurck formalism is to determine the most general \emph{ansatz} for the metric compatible with the desired symmetries. We are interested in static and axisymmetric solutions. This means a hypersurface orthogonal Killing field $k\equiv \partial/\partial t$ exists and is everywhere timelike as well as a stationary Killing field $m\equiv \partial /\partial \phi$, whose orbits are isomorphic to $U(1)$. Using $t$ and $\phi$ as coordinates leads to the following general line element
\begin{equation}
\mathrm{d}s^2=G_{tt}(r,\theta) \mathrm{d}t^2+G_{rr}(r,\theta)\mathrm{d}r^2+2 G_{r\theta}(r,\theta)\mathrm{d}r\,\mathrm{d}\theta+G_{\theta\theta}(r,\theta)\mathrm{d}\theta^2+G_{\phi\phi}(r,\theta) \mathrm{d}\phi^2\,,
\end{equation}
which still exhibits full reparameterization invariance in the $(r,\theta)$ coordinates. The remaining gauge freedom is fixed via an appropriate choice of the so called reference metric, which is an essential ingredient of the DeTurck formalism.  One next imposes   the  Einstein-DeTurck equation:
\begin{equation}
R_{ab}+\frac{3}{L^2}g_{ab}-\nabla_{(a}\xi_{b)}=0\,,
\label{eq:deturk}
\end{equation}
where $\xi^a = g^{cd}\left[\Gamma^a_{cd}(g)-\Gamma^a_{cd}(\bar{g})\right]$, $\Gamma^a_{cd}(\mathfrak{g})$ is the Levi-Civita connection associated with a metric $\mathfrak{g}$, and $\bar{g}$ is the so called reference metric. The reference metric is chosen to have the same axis location as the metric we seek to find, and satisfies the same Dirichlet boundary conditions as $g$. Solutions of Eq.~(\ref{eq:deturk}) do not necessarily coincide with solutions of the Einstein equation, except if $\xi^a=0$. However, it turns out that this is always the case for stationary solutions enjoying the so called $(t,\phi)$ symmetry \cite{Figueras:2016nmo}, and thus, in particular, for the static solutions we wish to determine (for the static case an alternative proof for the equivalence of solutions of the Einstein and Einstein-DeTurck equations was presented in \cite{Figueras:2011va}).

The next step is an appropriate choice of reference metric. We motivate it by starting in global AdS in standard $(t,r,\theta,\phi)$ coordinates and by preforming the following change of variables
\begin{equation}
r=L\,\frac{y\sqrt{2-y^2}}{1-y^2}\,,\quad\ \cos\theta = x\sqrt{2-x^2}\,,\quad \text{and}\quad t = L T\,,
\label{eq:trans}
\end{equation}
which transforms
\begin{equation}
\mathrm{d}s^2=-\left(1+\frac{r^2}{L^2}\right)\mathrm{d}t^2+\frac{\mathrm{d}r^2}{\displaystyle 1+\frac{r^2}{L^2}}+r^2\left(\mathrm{d}\theta^2+\sin^2\theta\mathrm{d}\phi^2\right)
\end{equation}
into
\begin{equation}
\mathrm{d}s^2=\frac{L^2}{(1-y^2)}\left\{-\mathrm{d}T^2+\frac{4\,\mathrm{d}y^2}{2-y^2}+y^2(2-y^2)\left[\frac{4\,\mathrm{d}x^2}{2-x^2}+(1-x^2)^2\mathrm{d}\phi^2\right]\right\}\,.
\end{equation}

For the reference metric we then take
\begin{subequations}
\begin{equation}
\mathrm{d}\bar{s}^2=\frac{L^2}{(1-y^2)}\left\{-\mathrm{d}T^2+\frac{4\,\mathrm{d}y^2}{2-y^2}+y^2(2-y^2)H(x,y)\left[\frac{4\,\mathrm{d}x^2}{2-x^2}+(1-x^2)^2\mathrm{d}\phi^2\right]\right\}\,,
\label{eq:ref}
\end{equation}
where
\begin{equation}
H(x,y)=(1-y^2)^2+y^2(2-y^2)h(x)\,,\qquad\text{with}\qquad h(x)=1+\epsilon\,\mathbb{S}_{\ell}\left(x\sqrt{2-x^2}\right)\,.
\end{equation}
\end{subequations}
Note that $H(x,0)=1$ and $H(x,1)=h(x)$. Our metric \emph{ansatz} now reads
\begin{multline}
\mathrm{d}s^2=\frac{L^2}{(1-y^2)}\Bigg\{-q_1(x,y)\,\mathrm{d}T^2+\frac{4\,q_2(x,y)}{2-y^2}\left(\mathrm{d}y+y\,q_3(x,y)\,\mathrm{d}x\right)^2+
\\y^2(2-y^2)H(x,y)\left[\frac{4\,q_4(x,y)\,\mathrm{d}x^2}{2-x^2}+(1-x^2)^2\,q_5(x,y)\,\mathrm{d}\phi^2\right]\Bigg\}\,,
\label{eq:ansatzdeturck}
\end{multline}
where $q_i(x,y)$, $i\in\{1,\dots,5\}$ are functions of $x$ and $y$ to be determined next. We will be interested in considering even values of $\ell$ only, for which $\mathbb{S}_{\ell}(\theta)$ is even under the reflection $\theta\to\pi-\theta$. This translates into parity symmetry under the transformation $x\to-x$, which we can use to half our computation domain. So we take $(x,y)\in[0,1]^2$, with $x=0$ being a symmetry hyperplane, $y=0$ the regular centre, $y=1$ the location of the conformal boundary and $x=1$ the north pole of the deformed $S^2$.

Next we address the issue of boundary conditions. At $y=0$, our geometry should locally look like flat space, because this corresponds to a point. This in turn enforces
\begin{equation}
q_4(x,0)=q_5(x,0)=q_2(x,0)\qquad\text{and}\qquad \left.\frac{\partial q_i(x,y)}{\partial y}\right|_{y=0}=0\,.
\end{equation}
At $x=0$ reflection symmetry imposes
\begin{equation}
\left.\frac{\partial q_1(x,y)}{\partial x}\right|_{x=0}=\left.\frac{\partial q_2(x,y)}{\partial x}\right|_{x=0}=\left.\frac{\partial q_4(x,y)}{\partial x}\right|_{x=0}=\left.\frac{\partial q_5(x,y)}{\partial x}\right|_{x=0}=0\qquad \text{and}\quad q_3(0,y)=0\,.
\end{equation}
At $x=1$, the north pole of the deformed two sphere, smoothness imposes
\begin{align}
&\left.\frac{\partial q_1(x,y)}{\partial x}\right|_{x=1}=\left.\frac{\partial q_2(x,y)}{\partial x}\right|_{x=1}=\left.\frac{\partial q_4(x,y)}{\partial x}\right|_{x=1}=\left.\frac{\partial q_5(x,y)}{\partial x}\right|_{x=1}=0\nonumber
\\
&\nonumber
\\
&q_3(1,y)=0\,,\quad \text{and}\quad q_4(1,y)=q_5(1,y)\,.
\end{align}

Finally, at the conformal boundary, we take $q_1(x,1)=q_2(x,1)=q_4(x,1)=q_5(x,1)=1$ and $q_3(x,1)=0$, which means that the line element (\ref{eq:ansatzdeturck}) is chosen to approach the reference metric (\ref{eq:ref}). All we need to show is that the reference metric (\ref{eq:ref}) has the correct boundary metric. This is easily seen if we use the coordinate transformation (\ref{eq:trans}).

To solve the Einstein-DeTruck equation numerically, we discretise the equations using a Chebyshev grid on Gauss-Lobatto nodes and solve the resulting equations using a standard Newton-Raphson routine. Such methods have been reviewed extensively in \cite{Dias:2015nua}.
\subsection{Results and comparisons}
\begin{figure}[ht]
\begin{center}
\includegraphics[width=.45\textwidth]{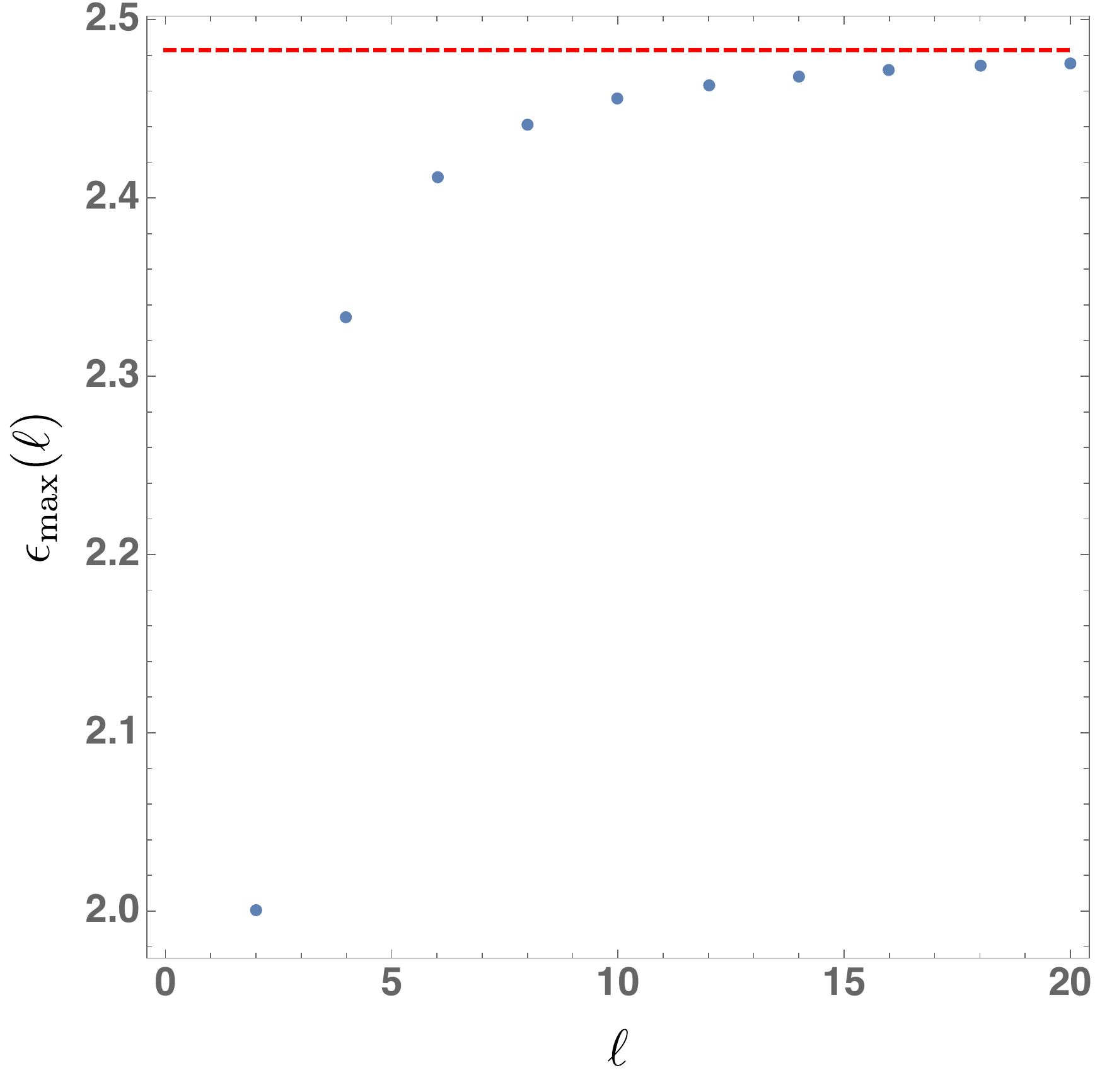}
 \caption{\label{fig:emax}$\epsilon_{\max}$ as a function of $\ell$: the blue disks are the exact numerical data and the red dashed curve the leading approximation at large $\ell$.}
 \end{center}
\end{figure}

Having constructed the new vacuum solutions, we now ask whether there are stable, static solutions for particles attached to cosmic strings with the same energy as the unbroken string.
We first note that the deformation parameter, $\epsilon$, is necessarily bounded above, since the boundary metric would otherwise have a curvature singularity (this is best illustrated by computing the Ricci scalar of the boundary metric) at the location of the zero of $1+\epsilon\,\mathbb{S}_{\ell}(\theta)$. In Fig.~\ref{fig:emax} we plot $\epsilon_{\max}(\ell)$ as a function of $\ell$ up to $\ell=20$. It is easy to show that, asymptotically, $\epsilon_{\max}\approx-\frac{1}{J_0\left(j_{1,1}\right)}+\mathcal{O}(1/\ell)$, where $J_k(x)$ is a $J$ Bessel function of order $k$, and $j_{k,l}$ is the $l^{\mathrm{th}}$ zero of the $J$ Bessel function of order $k$.

\begin{figure}[ht]
\begin{center}
\includegraphics[width=.45\textwidth]{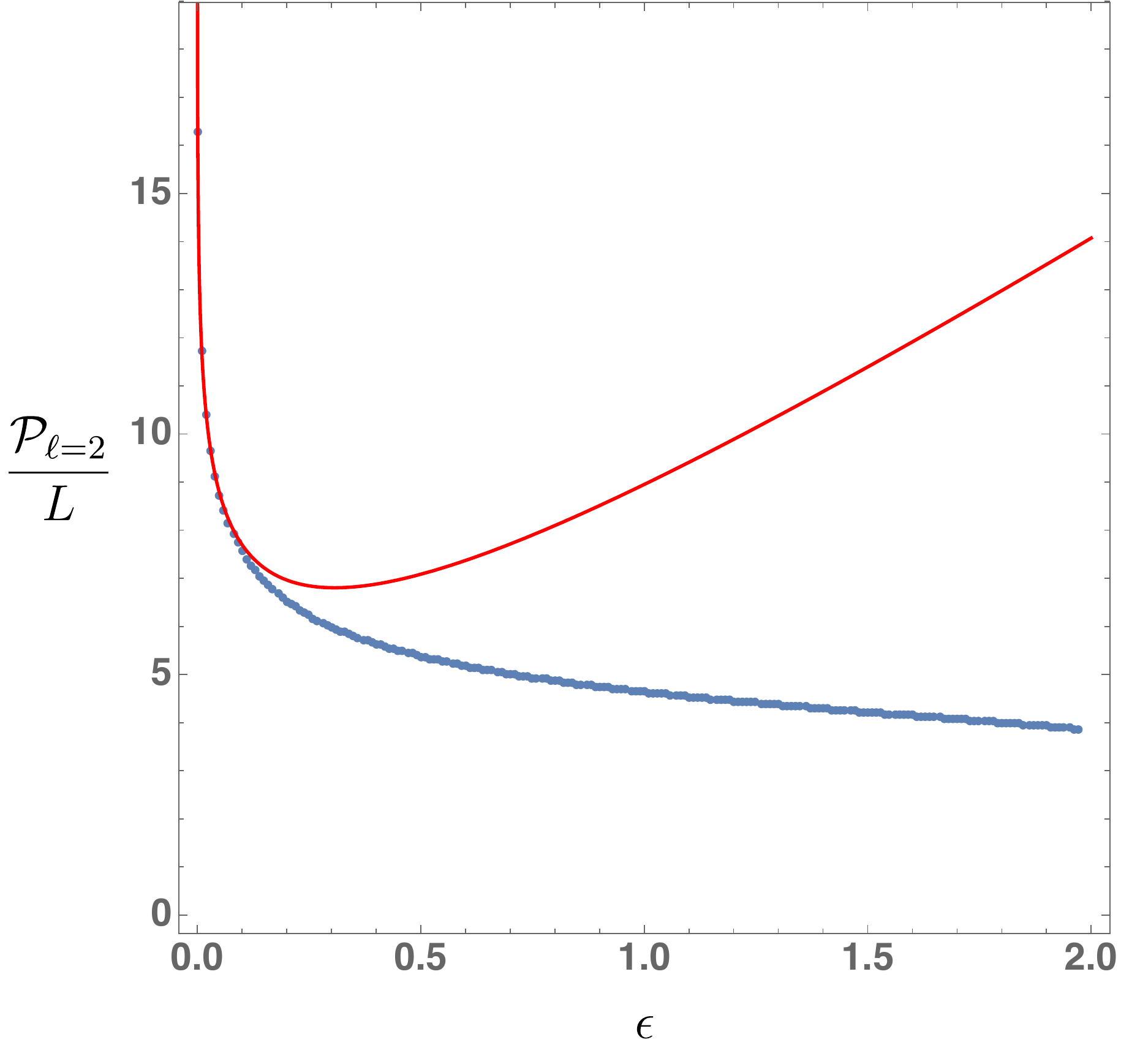}
 \caption{\label{fig:proper2}Proper distance between the two stable orbits computed for $\ell=2$: the blue disks correspond to our exact numerical data and the solid red line to the perturbative expression (\ref{eq:properpredi}).}
 \end{center}
\end{figure}

We begin by studying the case with $\ell=2$, for which $\epsilon_{\max}=2$, and briefly mention at the end what happens for different even values of $\ell>2$. First, we would like to understand whether stable orbits exist for arbitrarily small values of $\epsilon$, as predicted by our perturbative non-linear results. This indeed turns out to be the case. In Fig.~\ref{fig:proper2} we compute the proper distance between the two orbits as a function of $\epsilon$ for $\ell=2$. The blue disks correspond to our exact numerical results and the solid red line to the analytic expression (\ref{eq:properpredi}): the agreement between the two when $\epsilon\ll1$ is reassuring.
As we increase the deformation on the boundary, the separation between the static particles decreases, but always stays larger than the AdS radius.

\begin{figure}[ht]
\begin{center}
\includegraphics[width=.45\textwidth]{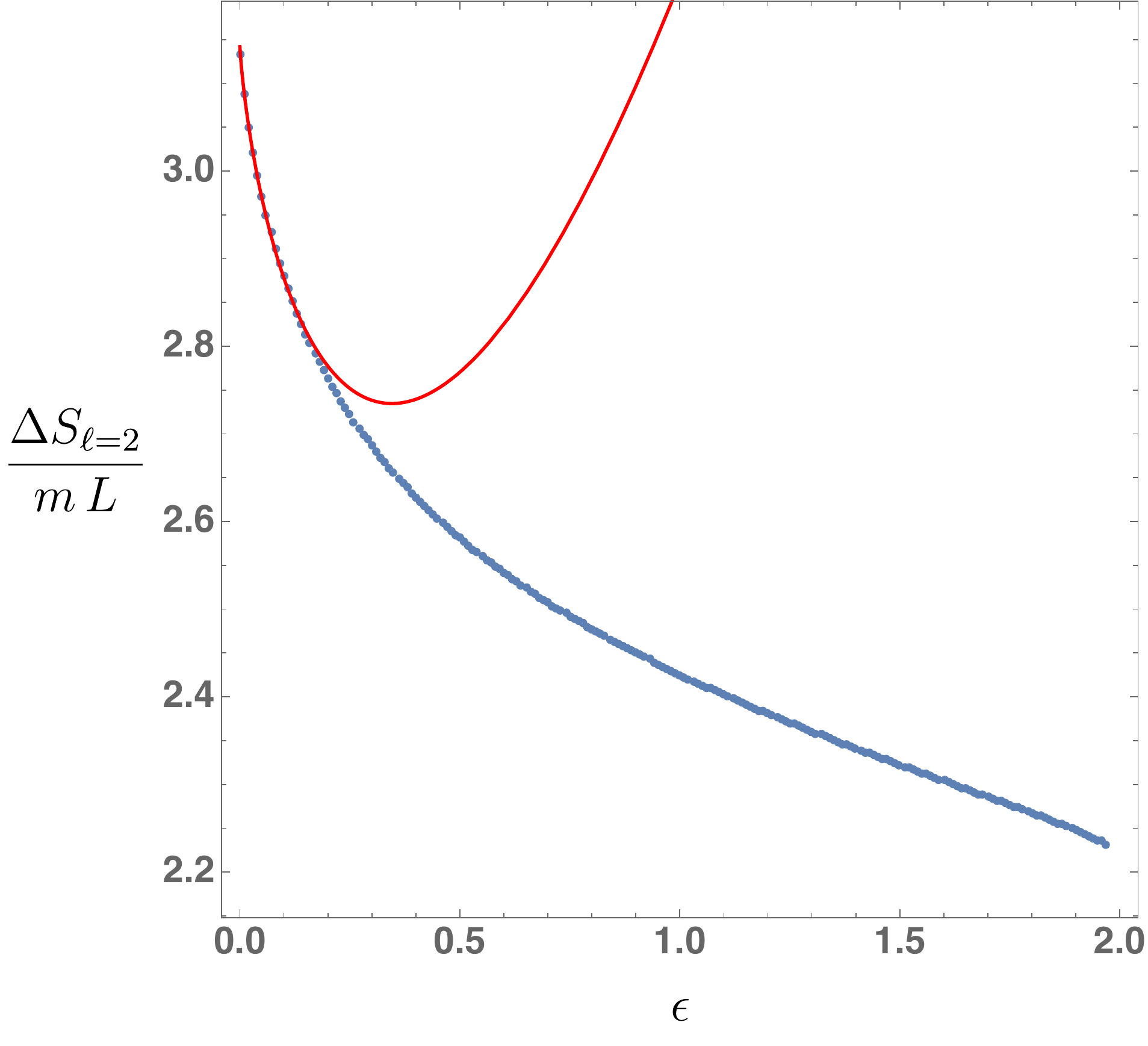}
 \caption{\label{fig:prod2}Instanton action computed for the $\ell=2$ profile: the blue disks correspond to our exact numerical data and the solid red line to the perturbative expression (\ref{eq:actionprodpre}).}
 \end{center}
\end{figure}

Next we compute the instanton action by computing (\ref{eq:actionproduction}) using our numerical solutions. The result of this calculation is seen in Fig.~\ref{fig:prod2}, where our numerical data is represented by the blue disks and the solid red line corresponds to our perturbative result (\ref{eq:actionprodpre}). Again, the agreement between these two expressions at small values of $\epsilon$ is indicative of the correctness of both methods. As expected, the action decreases as the separation between the particles decreases. Note that the nonlinear perturbative solution and the exact numerical solution are obtained in different gauges.

\begin{figure}[ht]
\begin{center}
\includegraphics[width=.95\textwidth]{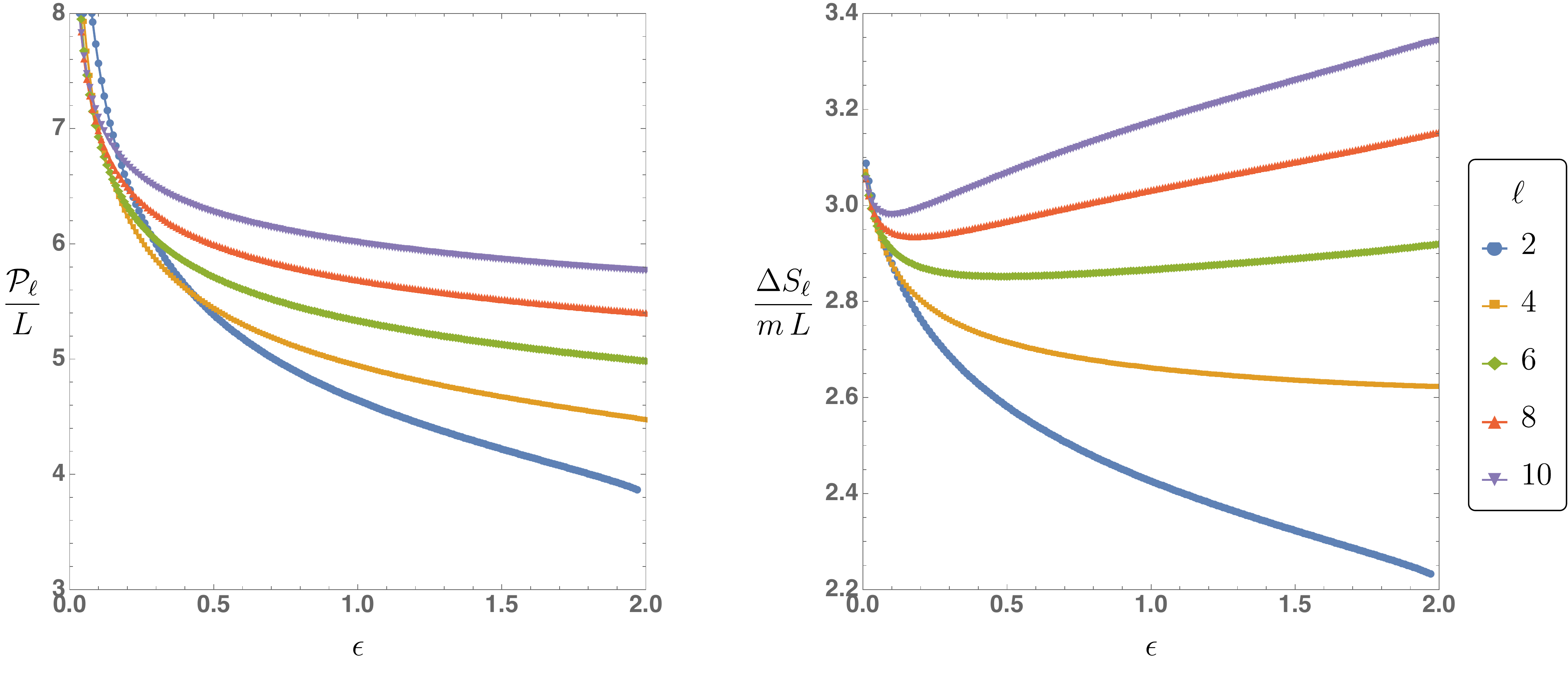}
 \caption{\label{fig:properactionell}Proper distance between orbits (left panel) and instanton production rate (right panel) computed for $\ell=2,4,6,8,10$ plotted as a function of $\epsilon$, with each curve labelling a different value of $\ell$.}
 \end{center}
\end{figure}

We have extended our calculations to other even values of $\ell$, and find that increasing $\ell$ increases $\mathcal{P}_{\ell}$ for large $\epsilon$ and decreases $\mathcal{P}_{\ell}$ for small enough $\epsilon$. We can observe this effect on the left panel of Fig.~\ref{fig:properactionell} where we plot the proper distance between orbits for several values of $\ell$, each of which are labelled in the figure caption. The behaviour at small $\epsilon$ is corroborated by the nonlinear perturbative prediction (\ref{eq:prediPell}). Nevertheless, for each value of $\ell$, we observe that increasing $\epsilon$ does decrease $\mathcal{P}_{\ell}$.

The behaviour of the instanton action  for various $\ell$ is shown  on the right panel of Fig.~\ref{fig:properactionell}. The action increases with $\ell$, and unlike the $\ell = 2$ case, for sufficiently large $\ell$ the action grows with $\epsilon$.

\section{Discussion}
We have presented examples of spacetimes in which a (test) cosmic string can break and produce two particles with arbitrarily small acceleration. There is a long history of showing that small black holes follow geodesics (see \cite{Gralla:2008fg} for a fairly recent discussion and references to earlier work). Similar arguments should show that particles attached to cosmic strings can be replaced by small black holes with no change in their dynamics. (The Euclidean solutions may have conical singularities on the horizon but they still describe pair creation \cite{Bousso:1998na}.)

Previous examples of black hole pair creation show that the black holes are created with their horizons identified at the moment of creation \cite{Garfinkle:1990eq} and therefore that the instanton creates a wormhole. In the limiting case of zero acceleration and for instantons free of conical singularities, the path integral computing the wormhole's quantum state is precisely of the form studied in \cite{Fu:2018oaq}, so with appropriate boundary conditions the back-reaction of quantum fields will render the wormhole traversable in the semiclassical approximation.  (We again remind the reader that some large parameter is generally required to render quantum fluctuations small \cite{Maldacena:2018gjk}.)  As argued in the introduction, the same will be true with sufficiently small non-zero acceleration; see in particular \eqref{eq:cMwh} for the condition relevant to the traversable wormholes of \cite{Maldacena:2018gjk}.

The net result is the creation of a traversable wormhole by a nonperturbative process in quantum gravity. We have seen that while this is not possible using only light cosmic strings in pure AdS, it is possible in nearby vacuum solutions.  It is interesting to note that these solutions  all have negative total energy  \cite{Hickling:2015tza}. This leads to the following intuitive picture: the attractive force of pure AdS is not quite sufficient to create the desired trajectories, but when we deform the boundary metric we lower the energy, effectively producing a deeper potential well in which it is possible.

When we replace the test particle with a small black hole, we only change the Euclidean solution  in a small tubular neighborhood of the particle trajectory which has  topology $S^1 \times {\mathbb R}^3$. The $S^1$ represents the Euclidean time along the particle worldline. To have smooth instantons, this period must match the inverse temperature of the black hole.  Since the particles in our instantons travel a long Euclidean proper distance before their orbits close, we can find such smooth black hole instantons by pasting in nearly-extremal Reissner-Nordstr\"om or Kerr black holes.  If the right sort of fields propagate in the wormhole throat, this should result in a wormhole that is traversable for long periods of time.  The perturbative argument is given in \cite{Fu:2018oaq}, and it is also expected from continuity with the non-perturbative constructions of \cite{Maldacena:2018lmt,Maldacena:2018gjk} which are related to thermofield double states at extremely small temperatures.  If the acceleration is small enough, we may expect the instanton to produce wormholes that remain open forever.

However, one should recall that instantons with conical singularities can sometimes yield higher pair-creation rates than smooth ones (see e.g. \cite{Bousso:1998na}).  This seems likely to occur in our context, as black holes farther from extremality will have greater entropy, so reducing the charge and angular momentum at fixed mass should increase the rate of pair production\footnote{In general, smooth instantons will extremize the action subject to varying the black hole area while holding fixed other parameters and boundary conditions; see e.g. the recent discussion in \cite{Dong:2018seb}.  But in \cite{Bousso:1998na} this extremum has a negative mode and is not a minimum.  In addition, in our case we remove the singularity by adjusting not only the black hole area, but also both the charge $q$ and associated potentials at infinity while holding fixed the location of the trajectory.  Extremizing the action under such variations need not remove conical singularities.}.  If one seeks a traversable wormhole that remains open for long times, it may thus not suffice to simply wait for the first wormhole to appear on our static trajectory.  One must also wait some additional time for one that is nearly extremal.  On the other hand, the fact that energy conservation fixes the mass $m$ of any particle created by our instanton near the static orbit leads to a positive feature of our process: one may choose parameters (including the string tension $\mu$) to make this $m$ large, perhaps even larger than our current approximations justify,  and then have the comfort of knowing that only large wormholes will be created at that location\footnote{Replacing our cosmic string with a magnetic flux tube should lead to an interesting variation.  At first order in the magnetic charge $q$, the mass of particles created by decay of the magnetic field at a given separation is proportional to $q$.  So watching a fixed static orbit selects a given charge-to-mass ratio $q/m$ rather than the fixed mass $m$ obtained in our case.  One should thus be able to engineer an instanton that creates {\it only} nearly-extreme black hole pairs regardless of whether conical singularities are allowed at the horizon.  But then the dominant wormholes created on the static orbit are likely to be small and large ones will be more rare.}.

It is not yet clear how large is the class of spacetimes which allow the creation of a traversable wormholes. In the vacuum examples discussed in the previous section, the particles are always created with a separation larger than the AdS radius $L$. We have also tried many other deformations of the boundary metric to see if we could lower the distance between the two particles below $L$, but were unable to do so. This suggests that the AdS attraction is playing a key role and similar phenomena may not be possible in vacuum asymptotically flat spacetimes.

 In addition to the examples discussed here, we have also found further examples by deforming the AdS soliton \cite{Horowitz:1998ha}. However, in that case there was a minimum deformation required to find stable static orbits. Also in that case, we were never able to decrease the distance between the two orbits to be below the AdS length scale $L$.

We also tried a number of asymptotically AdS spacetimes for which it does not seem to be possible to nucleate a traversable wormhole.   We have constructed (perturbatively) a number of boson stars in AdS for a wide range of mass parameter, with alternative and standard quantization, and we were not able to find stable orbits. We also constructed smooth geometries in global AdS with scalars satisfying double trace boundary conditions, and for those, also no orbits were found.

\section*{Acknowledgements}

G.H., D.M., and D.W. were supported in part by NSF grant PHY1801805. J.E.S. is supported in part by STFC grants PHY-1504541 and ST/P000681/1. This work used the DIRAC Shared Memory Processing system at the University of Cambridge, operated by the COSMOS Project at the Department of Applied Mathematics and Theoretical Physics on behalf of the STFC DiRAC HPC Facility ({www.dirac.ac.uk}). This equipment was funded by BIS National E-infrastructure capital grant ST/J005673/1, STFC capital grant ST/H008586/1, and STFC DiRAC Operations grant ST/K00333X/1. DiRAC is part of the National e-Infrastructure.

\bibliographystyle{JHEP}
\bibliography{library}

\end{document}